\documentclass[12pt,a4paper]{article} 
\usepackage{graphicx}
\usepackage[latin1]{inputenc} 
\usepackage[T1]{fontenc} 
\usepackage{amsmath}
\usepackage{amsfonts} 
\usepackage{amssymb} 
\usepackage[english]{babel}
\title{Vacuum energy via dimensional reduction of functional 
determinants}
\author{C.~D.~Fosco
and
F.~D.~ Mazzitelli\\
{\normalsize\it Centro At\'omico Bariloche}\\
{\normalsize\it Comisi\'on Nacional de Energ\'\i a At\'omica}\\
{\normalsize\it R8402AGP Bariloche, Argentina}\\
{\normalsize\it and}\\
{\normalsize\it Instituto Balseiro}\\
{\normalsize\it Universidad Nacional de Cuyo}\\
{\normalsize\it R8402AGP Bariloche, Argentina.}}

\begin{document} 
\date{}
\maketitle
\begin{abstract} 
\noindent 
We apply a `dimensional reduction' mechanism to the evaluation of the
functional integral for the vacuum energy of a real scalar
field in the presence of non-trivial backgrounds, in $d+1$ dimensions. The reduction is
implemented by applying a generalized version of Gelfand-Yaglom's theorem
to the corresponding functional determinant.
The main outcome of that procedure is an alternative representation for the
Casimir energy, which involves one spatial dimension less than the original
problem. We show that, for some configurations, important information about
the reduced problem can be obtained. We also show that the reduced problem
allows for the introduction of an approximation scheme which is novel
within this context.
\end{abstract}
\section{Introduction}\label{sec:intro}
Fluctuation determinants are ubiquitous objects in the quantum field theory
realm~\cite{Dunne:2007rt}, specially when using the functional integral
approach~\cite{ZinnJustin:2002ru}. Those determinants arise, for example,
when evaluating the effective action to the one-loop order, in particular
when one allows for a non-trivial background.  A well-known example
corresponds to the calculation of (one-loop) quantum corrections on top of
a soliton background~\cite{Rajaraman:1982is,Bordag:1994jz}.

Yet another example, somewhat related to the latter, is that of the Casimir
effect \cite{booksCasimir}. Here, the non-trivial background represents both the
geometry and properties of the mirrors, as extensively studied
in~\cite{Graham:2002fw}.  
In a previous paper~\cite{CcapaTtira:2011ga}, we considered precisely that kind
of situation: within the path integral approach, we applied the
Gelfand-Yaglom~\cite{Gelfand:1959nq,Kirsten:2004qv,Dunne:2007rt} (G-Y)
theorem to evaluate the corresponding functional determinant. This
provides an alternative derivation of Lifshitz formula \cite{Lifshitz:1956zz} in $d+1$ spacetime
dimensions, for the geometry where that formula applies. Namely,
translation symmetry is broken along just one direction, since the
mirrors are assumed to be flat, parallel, and composed of homogeneous media.

In this paper, we have in mind a more general configuration for this kind of problem,
namely, still a real scalar field in $d+1$ dimensions, but now in the presence of
mirrors (modeled by potentials)  which may have a more general shape.  

A common feature with the approach of~\cite{CcapaTtira:2011ga} is that, by applying 
a generalized version of the G-Y theorem, the corresponding functional determinant 
is converted into an equivalent expression; this time it amounts to a
lower-dimensional one. As we will see, in some cases, the latter may be
solved, even for geometries that differ from the one in Lifshitz formula. 
Besides, the very process used to reduce the dimensionality of the
problem will allow us to introduce novel approximation schemes, suitable
when some extra assumptions are made about the mirrors.

This paper is organized as follows: in Sect.~\ref{sec:casi}, we formulate
and introduce the notations and conventions about the problem we deal with
in this article, namely, about the Casimir energy in terms of a functional
determinant. Then, the dimensional reduction itself is presented in
Sect.~\ref{sec:redux}. In Sect.~\ref{sec:app} we consider different
applications of the results presented in Sect.~\ref{sec:redux}.
Finally, in Sect.~\ref{sec:concu} we present our conclusions. 

\section{Vacuum energy}\label{sec:casi}
The system that we consider involves a real scalar field $\varphi(x)$  in
$d+1$ Euclidean dimensions, with spacetime coordinates denoted by $x =
(x_0,\ldots,x_d)$, such that its action ${\mathcal S}(\varphi)$ has the form:
\begin{equation}
{\mathcal S}(\varphi) \;=\; \frac{1}{2} \, \int d^{d+1}x \, 
\left[ \big( \partial \varphi (x) \big)^2 + V({\mathbf x}) \big( \varphi(x)
\big)^2 \right]  \;,
\end{equation}
where ${\mathbf x}=(x_1,\ldots,x_d)$ denotes the spatial coordinates, and the 
`potential' $V$ accounts for the presence of the (static) mirrors. In terms
of ${\mathcal S}(\varphi)$, ${\mathcal Z}$, the Euclidean vacuum
persistence amplitude, is given by the path integral:
\begin{equation}
{\mathcal Z} \;=\; \int {\mathcal D}\varphi \; e^{- {\mathcal S}(\varphi)}
\;.
\end{equation} 
We assume the system to be defined  within a `space-time box':
$-\frac{L_i}{2} < x_i < \frac{L_i}{2}$ ($i =1, \ldots, d$), $-\frac{T}{2} <
\tau < \frac{T}{2}$ with, in principle, Dirichlet boundary conditions on
all the borders. Time translation invariance is broken, but
will be recovered when letting $T$ go to infinity at the end, a necessary step
to extract the {\em vacuum\/} expectation values of observables.  
At the $T \to \infty$ limit, the specific choice of boundary conditions at the
initial and final times has no bearing on the results.  Regarding the spatial 
borders of the box, although the boundary conditions are, as we said, assumed to
be of the Dirichlet type, we shall briefly comment on the differences which 
arise when Neumann conditions are imposed for one of the directions. 

Let us consider now the large-$T$ limit: since the  potential is
time independent, the leading behaviour of the effective action $\Gamma$ is 
$\Gamma \, \sim \, T \times E$, $E$ being the vacuum energy. Thus we may
relate $E$ to a functional determinant, by using the formal result for 
$\Gamma$ which, ignoring irrelevant factors, is given by:
\begin{equation}\label{eq:defdet}
	e^{-\Gamma} \;=\;  \big[\det(-\partial^2 + V
	\big)\big]^{-\frac{1}{2}} \;,
\end{equation}
with $\partial^2 \equiv \partial_\mu \partial_\mu$,  $\mu = 0,1,\ldots,d$.

Thus,
\begin{equation}\label{eq:gen}
	E\;=\; \frac{1}{2} \, \int_{-\infty}^{+\infty} \frac{dk_0}{2\pi} \; 
	\log \big(\det{\mathbb K}\big) \;,
\end{equation}
where
\begin{equation}\label{eq:defkgen}
	{\mathbb K}\;=\; -\nabla_{\mathbf x}^2 + V({\mathbf x}) \,+\, k_0^2 \;.
\end{equation}
Eq.(\ref{eq:gen}) yields the energy in the general case, i.e., without
making further simplifying assumptions about the potential. 

When the potential is invariant under a continuous spatial symmetry group $G$, the energy becomes proportional to $V_G$ the volume of $G$. Thus, when there is symmetry under translations, since the group's volume diverges, one usually introduces the corresponding energy per unit area, ${\mathcal E}$. For example, assuming invariance under translations in all the coordinates but $x_1$ and $x_2$, 
\begin{equation}
	{\mathcal E} \;=\; \lim_{T, L_3, \ldots, L_d \to
	\infty} \frac{\Gamma}{T L_3 \ldots L_d} \;,
\end{equation}
and,
\begin{equation}\label{eq:cale}
	{\mathcal E}\;=\; \frac{1}{2} \, \int \frac{d^dk_\parallel}{(2\pi)^{d-2}} 
	\log \big(\det{\mathbb K}'\big) \;,
\end{equation}
where 
\begin{equation}\label{eq:defkpart}
	{\mathbb K}'\;=\; - (\partial_1^2 + \partial_2^2) + V(x_1,x_2)
	\,+\, k_\parallel^2 \;,
\end{equation}
with $k_\parallel \equiv (k_0,k_3,\ldots,k_d)$.

\section{Dimensional reduction of the fluctuation
determinant}\label{sec:redux}

In this section, we perform a `dimensional reduction' of the functional
determinant of the operator ${\mathbb K}$ above.  By that we mean to
transform that to the determinant of a lower-dimensional operator.
Since, to do this, we apply G-Y theorem, let us  briefly
summarize it here, in a
fashion which best suits our needs. To that end, we have found it useful to
apply the theorem as presented in~\cite{Kleinert:2004ev}. 

To begin, let us assume that we want to evaluate the determinant of an $N \times N$
operator matrix ${\mathcal O} = [{\mathcal O}_{ij}]$, with $i,j,\ldots, N$,
acting on a space of $N$-component functions of a single variable $x$, 
satisfying Dirichlet conditions at $x = \pm \frac{L}{2}$, and defined by:
\begin{equation}
{\mathcal O}_{ij} \;=\; - \delta_{ij} \, \frac{d^2}{dx^2} \,+\,
\Omega_{ij}(x) \;,
\end{equation}
or, ${\mathcal O} =  -\frac{d^2}{dx^2} + \Omega(x)$.

To avoid irrelevant divergences, the result for $\det {\mathcal O}$ is usually written 
in terms of its ratio to the determinant of a {\em reference\/}
operator. For example, ${\mathcal O}_0$ could be defined as the
\begin{equation}
	{\mathcal O}_0 \;=\; ( -\frac{d^2}{dx^2} \,+\, \Omega_0 ) \delta_{ij} 
	\;,
\end{equation}
where $\Omega_0$ is a constant. 

Thus, the ratio between determinants, as given by the G-Y theorem, is:
\begin{equation}\label{eq:gy1}
\frac{\det{\mathcal O}}{\det{\mathcal O}_0} \;=\; 
\frac{\det D(x)}{\det D_0(x)}\Big|_{x=\frac{L}{2}} \;,
\end{equation}
where $D$ and $D_0$ denote $N \times N$ matrix functions of $x$, solutions to 
\begin{equation}
 [-\frac{d^2}{dx^2} + \Omega(x)] D(x) \;=\; 0 \;,\;\;
 [-\frac{d^2}{dx^2} + \Omega_0(x)] D_0(x) \;=\; 0 \;,
\end{equation}
with the initial conditions:
\begin{equation}
 [D_{ij}(x)]\Big|_{x=-\frac{L}{s}} \;=\; 0 \;\;,\;\;
 [\frac{d}{dx} D_{ij}(x)]\Big|_{x = -\frac{L}{2}} = \delta_{ij}  \;\;\;,
\end{equation}
and identical initial conditions for $D_0$. 

An important remark is that, in (\ref{eq:gy1}), the `det' symbol on the rhs refers 
just to the determinant of an $N \times N$ matrix, while in the lhs the
determinant is meant in its more general sense (i.e., including its functional
space action). Thus, in its usual set-up, the G-Y theorem provides a
dimensional reduction, from a functional determinant of functions depending
on one variable, to the determinant of an ordinary matrix.

Let us now turn to the evaluation of the functional determinant of the
operator ${\mathbb K}$ introduced above. The main idea is to single out one of the
coordinates, $x_d$, say, which will play the role of $x$ in
the G-Y theorem as presented above. The remaining coordinates and their
derivatives will be understood as defining an operator, the matrix elements
of which are identified with the ${\mathcal O}_{ij}$ ones. 

We first note that, without any loss of generality,
 ${\mathbb K}$ may be {\em interpreted\/}  as follows:
\begin{equation}\label{eq:ksplit}
	{\mathbb K} \;=\; -\partial_d^2 + {\mathbb H}(x_d) \;,
\end{equation}
where ${\mathbb H}(x_d)$ (which is, of course, the remainder of the
operator ${\mathbb K}$ after $-\partial_d^2$ has been subtracted) is also a
linear operator acting on functions of ${\mathbf x}$. 
However, it is convenient to split those coordinates as 
${\mathbf x} = ({\mathbf x}_\perp, x_d)$, where ${\mathbf x}$ denotes the
rest of the spatial coordinates:  $x_1,\ldots, x_{d-1}$. Then, ${\mathbb H}(x_d)$ may be
written as follows:
\begin{equation}
{\mathbb H}(x_d) \;\equiv\; - \nabla_\perp^2 \,+\, V_{x_d}({\mathbf x}_\perp) \,
+\, k_0^2 \;,
\end{equation}
with 
\begin{align}
	V_{x_d}({\mathbf x}_\perp) &\equiv V({\mathbf x}) \nonumber\\
	\nabla_\perp &\equiv \nabla_{{\mathbf x}_\perp} \;.
\end{align}	
The reason for introducing this seemingly formal splitting should be clear
from the form in which we have written the G-Y theorem;  in short, $x_d$ is
the variable about which the functional determinant of ${\mathbb K}$ will
be reduced.  We see that the operator in (\ref{eq:ksplit}) also acts on a
linear space of functions of $x_d$,  satisfying Dirichlet conditions on
$x_d = \pm \frac{L_d}{2}$. Note that, for each value of $x_d$, the would-be
discrete index $i$ in G-Y's theorem is here replaced by a dependence on the
${\mathbf x}_\perp$ variables, and the would be matrix elements are
operator kernels.  
Indeed, even though the G-Y theorem is usually formulated in terms of a
{\em finite\/} number of components, $N$, we recall that here, within the
path integral formulation, calculations are meant to be realized starting
from a discretized version of the continuous objects.  Thus, we keep the
continuous notation, but have in mind the previous statement, namely, that
the continuum is made sense of by reaching it from a finite dimensional
approximation, where the standard version of the G-Y theorem may be used.
To that end, we introduce the matrix elements with respect to the
coordinates ${\mathbf x}_\perp$, using Dirac's bra-ket notation, for
instance,
\begin{align}
\langle {\mathbf x}_\perp | {\mathbb K} | {\mathbf x}'_\perp \rangle 
&\;=\; K_{x_d}({\mathbf x}_\perp , {\mathbf x}'_\perp) \nonumber\\
&\;=\; -  \delta({\mathbf x}_\perp - {\mathbf x}'_{\perp}) \partial_d^2
\,+\, H_{x_d}({\mathbf x}_\perp , {\mathbf x}'_{\perp}) \;,
\end{align}
where
\begin{align}
& H_{x_d}({\mathbf x}_\perp , {\mathbf x}'_{\perp}) \,=\, \langle {\mathbf
x}_\perp | {\mathbb H}(x_d) | {\mathbf x}'_\perp \rangle \nonumber\\
&\;=\; - \nabla_{{\mathbf x}_\perp}^2 \delta({\mathbf x}_\perp - {\mathbf
x}'_{\perp})  \,+\, [ V_{x_d}({\mathbf x}_\perp) + k_0^2 ]  \, \delta({\mathbf
x}_\perp - {\mathbf x}'_{\perp}) \;.
\end{align}

The result for the determinant  is usually presented in terms of its ratio
with a `reference' operator ${\mathbb K}_0$. In the case at hand:
\begin{equation}\label{eq:gyf}
\frac{\det{\mathbb K}}{\det{\mathbb K}_0}\;=\;
\frac{\det \psi(\frac{L_d}{2})}{\det\psi_0(\frac{L_d}{2})} \;,
\end{equation}
where  $\psi(x_d)$ and $\psi_0(x_d)$ are solutions to the homogeneous equations
\begin{equation}\label{eq:hom1}
	{\mathbb K} \psi(x_d) \;=\; 0 \;\;,\;\;\;
	{\mathbb K}_0 \psi_0(x_d) \;=\; 0 \;.
\end{equation}
Note that they are also operators on the ${\mathbf x}_\perp$ coordinates,
namely, with matrix elements which we may naturally denote by:
\begin{equation}
\langle {\mathbf x}_\perp | \psi(x_d) | {\mathbf x}'_\perp \rangle \;=\;
\psi_{x_d}( {\mathbf x}_\perp , {\mathbf x}'_\perp ) \;,
\end{equation}
(and analogously for $\psi_0$). The initial conditions on the solutions to
the homogeneous equations are
\begin{align}
	& \big[ \psi_{x_d}( {\mathbf x}_\perp , {\mathbf x}'_\perp )
\big]\Big|_{x_d = - \frac{L_d}{2}} \;=\; 0 \nonumber\\
& \big[ \frac{\partial \psi_{x_d}}{\partial x_d}({\mathbf x}_\perp ,
{\mathbf x}'_\perp )
\big]\Big|_{x_d = - \frac{L_d}{2}} \;=\;  
\delta( {\mathbf x}_\perp - {\mathbf x}'_\perp ) \;.
\end{align}

Our next step is thus to write a more explicit solution to (\ref{eq:hom1}).
That can be done, by first converting the differential equation with
respect to $x_d$ to a first order one, by first introducing
\begin{equation}
	\Psi(x_d)\;\equiv\; 
	\left( 
		\begin{array}{c}
			\psi (x_d)\\
			\frac{\partial \psi}{\partial x_d}(x_d) 	
		\end{array}
	\right) \;,
\end{equation}
which renders the original second-order equation into 
\begin{equation}\label{eq:hom2}
	\frac{\partial \Psi}{\partial x_d}(x_d) \;=\; {\mathcal H}(x_d)
	\Psi(x_d) \;.
\end{equation}
with
\begin{equation}
{\mathcal H}(x_d) \;=\;	
	\left( 
		\begin{array}{cc}
			0 &  {\mathbb I}		\\
			{\mathbb H}(x_d) & 0 
		\end{array}
	\right) \;,
\end{equation}
where ${\mathbb I}$ denotes the identity operator.
Using an evolution operator ${\mathbb U}$, we then have the explicit
solution:
\begin{equation}
	\Psi(x_d) \;=\; {\mathbb U}(x_d,-L_d/2) \Psi(-L_d/2) 
\end{equation}	
with
\begin{equation}\label{eq:defu}
	{\mathbb U}(x''_d,x'_d) \,\equiv\,
	{\mathcal P} \exp\Big[ \int_{x'_d}^{x''_d} dy_d
	{\mathcal H}(y_d) \Big] \;\;\;\; (x''_d \geq x'_d) \;,
\end{equation}
where we have introduced the path-ordering operator ${\mathcal P}$, which acts 
in the same way as the time-ordering operator, but with the first spatial 
coordinate playing the role of the time.

Equipped with the solution for $\Psi$ just presented, we note that
\begin{equation}\label{eq:gyf1}
\left( \begin{array}{c}
\psi(L_d/2) \\
     0
\end{array} 
\right) \,=\, {\mathbb U}(L_d/2,-L_d/2) \left( \begin{array}{c}
    0 \\
{\mathbb I}
\end{array} 
\right)\;,
\end{equation}
or, using indices $A$ and $B$, which can assume the values $1$ or $2$, to 
distinguish the $4$ (operatorial) blocks in ${\mathbb U}(L_d/2,-L_d/2)$:
\begin{equation}\label{eq:mblocks}
{\mathbb U}(L_d/2,-L_d/2) \;\equiv\; 
\left( \begin{array}{cc}
{\mathbb U}_{11} & {\mathbb U}_{12} \\
{\mathbb U}_{21} & {\mathbb U}_{22} 
\end{array} 
\right)\;,
\end{equation}
we see that 
\begin{equation}
\frac{\det{\mathbb K}}{\det{\mathbb K}_0}\;=\; \frac{\det{\mathbb U}_{12}}{\det{\mathbb U}_{12}^{(0)}}\, .
\end{equation}
This is our main result. The original determinant in $d+1$ dimensions has been reduced to the computation of a determinant in one 
dimension less. Moreover, as we will see, the fact that ${\mathbb U}$ is an evolution operator will allow us to implement 
novel approximations to the evaluation of the determinant.

We then perform a consistency check which consists of verifying that the
result of the above procedure yields correct results in a well-known situation, 
namely, that of a potential which only depends on ${\mathbf x}_\perp$, i.e., 
$V_{x_d}({\mathbf x}_\perp) = V({\mathbf x}_\perp)$.  

We note that, under this assumption, there is no need to use a path-ordering 
to write ${\mathbb U}$, since ${\mathbb H}$ is independent of $x_d$ and 
therefore the operators ${\mathcal H}(x_d)$ commute for different values of $x_d$. Thus, 
taking into account that 
\begin{equation}
\left( \begin{array}{cc}
	 0 &  {\mathbb I}		\\
	{\mathbb H} & 0 
	\end{array}
\right)^{2 n} = \left( \begin{array}{cc}
	 {\mathbb H}^n &  0		\\
	0 & {\mathbb H}^n
	\end{array}\right)
	\, ,\quad
	\left( \begin{array}{cc}
	 0 &  {\mathbb I}		\\
	{\mathbb H} & 0 
	\end{array}
\right)^{2 n+1} = \left( \begin{array}{cc}
	0&  {\mathbb H}^{n} 	\\
	 {\mathbb H}^{n+1} & 0
	\end{array}\right)\, ,
	\end{equation}
we find
\begin{align}\label{eq:umat1}
{\mathbb U}(L_d/2,-L_d/2) \;=\; e^{L_d {\mathcal H}} \;& =\;
\frac{\sinh(L_d \sqrt{\mathbb H})}{\sqrt{\mathbb H}} \, 
\left( \begin{array}{cc}
	 0 &  {\mathbb I}		\\
	{\mathbb H} & 0 
	\end{array}
\right) \nonumber\\
\,&+\,\cosh(L_d \sqrt{\mathbb H}) \, 
\left( \begin{array}{cc}
		{\mathbb I}	& 0	\\
		0  & {\mathbb I}  
		\end{array}
\right) \;.
\end{align}
The (positive) square root of the seld-adjoint ${\mathbb H}$ operator have
been introduced in order to have a more compact expression. One can note, 
however, that the result is analytic in ${\mathbb H}$.

Thus, from (\ref{eq:gyf1}), 
\begin{equation}\label{eq:gyf2}
\frac{\det{\mathbb K}}{\det{\mathbb K}_0}
\;=\; \frac{\det[\sqrt{{\mathbb H}_0}]}{\det[\sqrt{\mathbb H}]}
\times \frac{\det[\sinh(L_d \sqrt{\mathbb H})]}{\det[\sinh(L_d \sqrt{{\mathbb H}_0})]}
\;,
\end{equation}
where, we recall, ${\mathbb H} = -\nabla^2_\perp + V({\mathbf
x}_\perp) + k_0^2$, 
and  ${\mathbb H}_0 = -\nabla^2_\perp + V_0 + k_0^2$. Note
that the determinants on the rhs of  Eq.(\ref{eq:gyf2}) correspond to operators
that act on functions of one coordinate less than the ones on the lhs. 

From the previous result Eq.(\ref{eq:gyf2}), we see that the Casimir energy $E$ 
can, in this case, be written as follows:
\begin{align}
E \;=\; &\frac{1}{2} \,\int_{-\infty}^{+\infty} \frac{dk_0}{2\pi}
\log\Big(\frac{\det{\mathbb H}_0}{\det {\mathbb H}}\Big)
\,+\, \frac{L_d}{2} \;\int_{-\infty}^{+\infty} \frac{dk_0}{2\pi} \,
{\rm Tr}\Big(\sqrt{\mathbb H} - \sqrt{{\mathbb H}_0}\Big)
\nonumber\\
\,&+\, \frac{1}{2} \;\int_{-\infty}^{+\infty} \frac{dk_0}{2\pi} \,
\log \frac{\det ({\mathbb I} - e^{- 2 L_d \sqrt{\mathbb H}})}{\det
({\mathbb I} - e^{- 2 L_d \sqrt{{\mathbb H}_0}})} \;,
\end{align}
which could be evaluated, in principle, from the knowledge of the
eigenvalues of ${\mathbb H}$.  Before introducing any model regarding the
form of the potential, we may note that, in the large $L_d$ limit, only the
second term in the previous equation is extensive in that length.  Thus,
the energy per unit length, when $L_d$ becomes infinite, is:
\begin{align}
{\mathcal E}\;&=\; \lim_{L_d\to \infty} \; \frac{E}{L_d} \nonumber\\
& =\; \frac{1}{2} \,\int_{-\infty}^{+\infty} \frac{dk_0}{2\pi} \, {\rm
Tr}\Big(\sqrt{\mathbb H} - \sqrt{{\mathbb H}_0}\Big) 
\end{align}
or, in terms of the eigenvalues  $(\lambda_n)^2$ and $(\lambda_n^{(0)})^2$ of
${\mathbb H}$ 
and ${\mathbb H}_0$,
\begin{equation}
{\mathcal E}\;=\; \frac{1}{2} \, \sum_n \big( \lambda_n \,-\,
\lambda_n^{(0)}\big) \;.
\end{equation}
Finally, to make contact with the usual expression for the Casimir energy
density, we introduce the eigenvalues $(\varepsilon_n)^2$ and
$(\varepsilon_n^{(0)})^2$ of $h \equiv -\nabla_\perp^2 +
V({\mathbf x}_\perp)$ and $h_0 \equiv -\nabla_\perp^2 +
V_0$, respectively,
\begin{equation}
	{\mathcal E}\;=\; \frac{1}{2} \, \int_{-\infty}^{+\infty}
	\frac{dk_0}{2\pi} \sum_n \big[ \sqrt{(\varepsilon_n)^2
+ k_0^2} - \sqrt{(\varepsilon_n^{(0)})^2 + k_0^2} \big]\;\;,
\end{equation}
which is the correct expression for the Casimir energy per unit length, in terms of the zero
point energies, when there is translation invariance along one direction.
It is interesting to note that, in the expression that we have obtained,
$k_0$ ends up playing the role of $k_d$, which is of course a dummy variable.

We can evaluate the energy $E$ exactly in this example, even for finite values of $L_d$, 
for some particular choices of ${\mathbb H}$, albeit the exact closed
expression is rather cumbersome, except, for example, when one considers
the classical Casimir effect limit. In this case, one evaluates the free
energy $F$. Let us assume, for example, that $d=2$ and 
that $V$ corresponds to an infinite potential well of width $L_1$ along the coordinate $x_1$:
\begin{align}
{\mathbb H} &=\; - \frac{d^2}{dx_1^2} + V(x_1)  \nonumber\\
V(x_1) &= \; \left\{ \begin{array}{ccc} 
0 & {\rm if} & 0 < x_1 < L_1 \\
\infty & {\rm if} & x_1 \leq 0 \;{\rm or}\; L_1 \leq x_1 
\end{array} 
\right. 
\end{align}

Neglecting $L_d$ and $L_1$-independent terms, we see that
\begin{equation}
F \;=\; \frac{1}{2 \beta} \, \log \eta( \tau) \;\;,\;\;\; \tau = i \frac{L_d}{L_1}\;,
\end{equation}
where $\eta(\tau)$ denotes Dedekind's $\eta$-function.

Another test one can perform starting from the  general expression for the fluctuation determinant corresponds to taking  the $L_d\to 0$ limit. In this case, that length becomes much shorter 
than all the other scales in the problem. In particular, than the scale of variation of the potential. Thus, we can use the expression corresponding to the result corresponding to a potential which is independent of $x_d$, with $V_{x_d}$ replaced by $V_{x_d=0}$:
 \begin{equation}\label{sinh}
\det(\mathbb K)\;\simeq\; \det \left[\frac{\sinh(L_d \sqrt{-\nabla^2_{\mathbf{x}_\perp} 
+ V_{x_d=0}({\mathbf x}_\perp) + k_0^2 })}{L_d \sqrt{-\nabla^2_{\mathbf{x}_\perp} 
+ V_{x_d=0}({\mathbf x}_\perp) + k_0^2 }} \right]
\;,
\end{equation}
where we have neglected $V$-dependent factors.

Thus, 
\begin{equation}\label{sinhapprox}
\det(\mathbb K)\;\simeq\; \det \left[ -\nabla^2_{\mathbf{x}_\perp} 
+ V_{x_d=0}({\mathbf x}_\perp) + k_0^2 + \frac{6}{L_d^2}\right]
\;,
\end{equation}
which is then the determinant one must consider, when implementing the
dimensional reduction for Dirichlet conditions on the borders of the small
dimension. Note that it is equivalent to the fluctuation determinant for a massive
real scalar field in $(d-1) +1 $ dimensions, with a
mass $\sqrt{6}/L_d$. To interpret this result in terms of a Kaluza-Klein
tower of massive modes, it is perhaps worth noting that, because of the use of
Dirichlet boundary conditions, all the modes are massive. Moreover,
the ratio between the different masses is independent of $L_d$.
Indeed, recalling the infinite-product expansion:
\begin{equation}
	\frac{\sinh x}{x} \;=\; \prod_{k=1}^\infty \big(1 + \frac{x^2}{\pi^2
	k^2} \big) \;,
\end{equation}	
we see that
\begin{equation}
	\det(\mathbb K)\;\simeq\; \prod_{l=1}^\infty  \det[-\nabla^2_{\mathbf{x}_\perp} 
+ V_{x_d=0}({\mathbf x}_\perp) + k_0^2 + m_l^2 ]  \;,
\end{equation}
where the mass $m_l$ of the $l$-mode is $\frac{l \pi}{L_d}$.
It is interesting to remark that the alternative expression we found  in Eq.\eqref{sinh},  allowed us to find a very simple expression, Eq.\eqref{sinhapprox},
for the  product of determinants associated to the tower of massive modes, in the $L_d\to 0$ limit.

We conclude this Section by considering the outcome of the same reduction
procedure, when Neumann boundary conditions (rather than Dirichlet ones)
are imposed along the $x_d$ direction. When Neumann conditions are used,
regarding the application of the G-Y theorem, one must use a different
initial condition for $\psi(x_d)$ (for the same homogeneous equation): 
\begin{align}
	& \big[ \psi_{x_d}( {\mathbf x}_\perp , {\mathbf x}'_\perp )
\big]\Big|_{x_d = - \frac{L_d}{2}} \;=\; \delta( {\mathbf x}_\perp - {\mathbf x}'_\perp ) \nonumber\\
& \big[ \frac{\partial \psi_{x_d}}{\partial x_d}({\mathbf x}_\perp ,
{\mathbf x}'_\perp )
\big]\Big|_{x_d = - \frac{L_d}{2}} \;=\;  0 \;,
\end{align}
and, besides, the determinant is given by the ratio between derivatives
along $x_d$ at the final point. Thus, the Neumann version of (\ref{eq:gyf}) is
\begin{equation}\label{eq:gyfn}
	\frac{[\det{\mathbb K}]_N}{[\det{\mathbb K}_0]_D}\;=\;
\frac{\det [\frac{\partial\psi}{\partial x_d}(\frac{L_d}{2})]}{\det 
[\frac{\partial\psi_0}{\partial x_d}(\frac{L_d}{2})]} \;,
\end{equation}
which of course may be extracted from the very same operator ${\mathbb U}$
introduced for the Dirichlet case: one just has to take different matrix
elements. In this vein, we may write the ratio between determinants
with either Dirichlet (D) or Neumann (N) conditions along $x_d$, as
follows:
\begin{equation}\label{eq:gyfdn}
\frac{\det{\mathbb K}_D}{\det{\mathbb K}_N}\;=\;
\frac{\det {\mathbb U}_{12}}{\det {\mathbb U}_{21}} \;.
\end{equation}
From (\ref{eq:umat1}) this implies, for the particular case of potentials
independent of $x_d$, that
\begin{equation}\label{eq:gyfdnf}
\frac{\det{\mathbb K}_N}{\det{\mathbb K}_D}\;=\; \det {\mathbb H} \;.
\end{equation}
Then we see that the difference between the corresponding vacuum energies
is:
\begin{equation}
 E_N - E_D \;=\; \frac{1}{2} \,\int_{-\infty}^{+\infty} 
 \frac{dk_0}{2\pi} \, \log \det {\mathbb H} \;,
\end{equation}
in other words, the difference may be identified as due to a
lower-dimensional fluctuation operator, which comes from a zero mode which
is only present in the Neumann case.

\section{Applications}\label{sec:app}
\subsection{Stepwise dependence of ${\mathbb H}(x_d)$ on $x_d$}
Let us consider here a situation such that the `Hamiltonian'~\footnote{We adopt 
the term Hamiltonian in view of the  similarity of ${\mathbb H}$ with the quantum 
mechanical Hamiltonian for a non-relativistic point particle in $d-1$ dimensions.} 
${\mathbb H}(x_d)$ may adopt one of two forms, ${\mathbb H}_a$ or ${\mathbb H}_b$, 
depending on whether $x_d$ is greater or smaller than $0$, respectively. 
Taking into account  that 
\begin{equation}
{\mathbb U}(L_d/2,-L_d/2)={\mathbb U}(L_d/2,0){\mathbb U}(0,-L_d/2)\, ,
\end{equation}
and the fact that in both factors the Hamiltonian does not depend on $x_d$, we can use a result similar to that  of Eq.\eqref{eq:umat1} 
for each one
to obtain
\begin{align}\label{eq:gyf3}
	\det{\mathbb K}
\;\propto\; & \det\Big[ \frac{\sinh(\frac{L_d}{2} \sqrt{\mathbb H}_a)}{\sqrt{\mathbb H}_a} 
\cosh(\frac{L_d}{2} \sqrt{\mathbb H}_b) 
\nonumber\\
\;&+ \;  \cosh(\frac{L_d}{2} \sqrt{\mathbb H}_a)  
\frac{\sinh(\frac{L_d}{2} \sqrt{\mathbb H}_b)}{\sqrt{\mathbb H}_b} \Big] 
\;.
\end{align}
where we have not yet introduced any reference operator, hence the
proportionality symbol. To fix that, we consider the
ratio with the determinant and that corresponding to a system which is uniform with
respect to the $x_d$ coordinate, with an operator ${\mathbb K}_0$, yet
unspecified: 
\begin{equation}\label{eq:gyf4}
\frac{\det{\mathbb K}}{\det{\mathbb K}_0}
\;=\; 
\frac{\det\Big[ \frac{\sinh(\frac{L_d}{2} \sqrt{\mathbb H}_a)}{\sqrt{\mathbb H}_a} 
\cosh(\frac{L_d}{2} \sqrt{\mathbb H}_b)  + \cosh(\frac{L_d}{2} \sqrt{\mathbb H}_a)  
\frac{\sinh(\frac{L_d}{2} \sqrt{\mathbb H}_b)}{\sqrt{\mathbb
H}_b}\Big]}{\det[ \frac{\sinh(L_d \sqrt{\mathbb H}_0)}{\sqrt{\mathbb H}_0}]} \;,
\end{equation}
where ${\mathbb H}_0$ corresponds to the reference operator ${\mathbb
K}_0$.

Taking the large-$L_d \to \infty$ limit, we get two kinds of contributions
to the {\em relative\/} vacuum energy $\delta E$, i.e., measured
with respect to the reference energy defined by ${\mathbb K}_0$: 
\begin{equation}
	\delta E \;\sim\; {\delta E}_{ex} \,+\, {\delta E}_{loc} \;,
\end{equation}
where ${\delta E}_{ex}$ is extensive in $L_d$,  while ${\delta E}_{loc}$ is
independent of that
length. The latter thus represents the effect of localized modes on
$x_d=0$, the place where there is a discontinuity in the functional 
operator. The explicit form of ${\delta E}_{ex}$ is
\begin{equation}
	{\delta E}_{ex} \;=\; L_d \, \big( \frac{{\mathcal E}_a \,+\,{\mathcal
	E}_b}{2}  - {\mathcal E}_0 \big) 
\end{equation}
where ${\mathcal E}_a$, ${\mathcal E}_b$, and ${\mathcal E}_0$ are the
vacuum energies per unit length corresponding to systems with
operators ${\mathbb H}_a$, ${\mathbb H}_b$, and ${\mathbb H}_0$,
respectively, over the whole range of the $x_d$ coordinate.   

The localized contribution is, on the other hand, given by:
\begin{equation}\label{eq:gyf5}
	{\delta E}_{loc} \,=\; E_{loc} \;=\; \frac{1}{2} \log \det
	\Big(\frac{1}{\sqrt{{\mathbb
	H}_a}} \,+\, \frac{1}{\sqrt{{\mathbb H}_b}} \Big) 
\;-\; \frac{1}{2} \log \det \Big(\frac{2}{\sqrt{{\mathbb H}_0}} \Big) \;,
\end{equation}
where the first equality follows from the fact that there is no localized
contribution from the reference opetator.

To proceed, naturally, we chose as reference an operator which exactly cancels the
extensive contribution. Such a choice is unique, and corresponds to: 
\begin{equation}
\sqrt{{\mathbb H}_0}  \;=\; \frac{\sqrt{{\mathbb H}_a} + \sqrt{{\mathbb
H}_b}}{2} \;,
\end{equation}
or:
\begin{equation}\label{eq:defh0}
	{\mathbb H}_0  \;=\; \frac{1}{2} \Big[ \frac{{\mathbb H}_a +
		{\mathbb H}_b}{2} \,+\, \frac{1}{2} \big(\sqrt{{\mathbb
		H}_a} \sqrt{{\mathbb H}_b}  + \sqrt{{\mathbb H}_b}
	\sqrt{{\mathbb H}_a} \big) \Big] \;. 
\end{equation}
Since this choice exactly cancels the vacuum energy with respect to that reference,
we can rephrase the previous fact by saying that {\em the extensive 
contribution to the vacuum energy is identical to the one of an $x_d$-translation
invariant system, equipped with a Hamiltonian defined by the equation
above\/}. 
Namely, $E_{ex}$ is the vacuum energy defined by ${\mathbb H}_0$. That means that the 
extensive contribution to the vacuum energy in a system like this can be calculated (perhaps
not unexpectedly) as the product of the length $L_d$ by the average of the two vacuum energies per
unit length, ${\mathcal E}_a$ and ${\mathcal E}_b$. 

At the same time, the choice renders the localized term in the following form:
\begin{equation}\label{eq:eloc}
	E_{loc} \;=\;  \frac{1}{2} \, \int_{-\infty}^{+\infty}
	\frac{dk_0}{2\pi} \; 
	\Big\{ \log(\det {\mathbb H}_0) - \frac{1}{2} \big[
	\log(\det {\mathbb H}_a) 
	+ \log(\det {\mathbb H}_b) \big] \Big\}\;, 
\end{equation}
which has the interpretation of being the difference between the vacuum energy
due to a lower dimensional system defined by ${\mathbb H}_0$ and the
average of the vacuum energies due to  two systems, defined in one less
dimension, with the Hamiltonians ${\mathbb H}_a$ and ${\mathbb H}_b$,
respectively.
This localized contribution, although precisely defined by (\ref{eq:eloc}), cannot
be easily calculated in a given non-trivial system. Indeed, in general it involves 
the determinant of ${\mathbb H}_0$, an object composed of two other Hamiltonians, 
${\mathbb H}_a$ and  ${\mathbb H}_b$, which generally do not commute.
 
 One can, however, find approximate expressions for $E_{loc}$ when some extra 
assumptions are made, namely, that the difference between ${\mathbb H}_a$ and ${\mathbb H}_b$ is small, 
in comparison with ${\mathbb H}_a$ (or ${\mathbb H}_b$). Setting:
\begin{equation}
{\mathbb H}_b \;=\; {\mathbb H}_a \,+\, {\mathbb H}' \;,
\end{equation}
we then expand $E_{loc}$ under the assumption of small ${\mathbb H}'$. The zeroth-order
term vanishes trivially, and one can verify, by explicit calculation, that so does the first-order 
one. 

After some algebra, the second-order term, $E_{loc}^{(2)}$, can be shown to be:
\begin{equation}\label{eq:eloc2}
E_{loc}^{(2)} \;=\; \frac{1}{32} \int_{-\infty}^{+\infty} \frac{dk_0}{2\pi} \, 
{\rm Tr}\left[ ({\mathbb H}_a)^{-1} {\mathbb H}' 
({\mathbb H}_a)^{-1} {\mathbb H}'\right] \;.
\end{equation} 
To proceed further, we assume that the two Hamiltonians differ in their 
potentials, namely: ${\mathbb H}' \,=\, \delta V({\mathbf x}_\perp)$, we see that 
$E_{loc}^{(2)}$ may be rendered in the form:
\begin{equation}\label{eq:eloc21}
E_{loc}^{(2)} \;=\; \frac{1}{2} \, \int d^2{\mathbf x}_\perp 
\int d^2{\mathbf x}'_\perp \, F({\mathbf x}_\perp ,{\mathbf x}'_\perp) \;
\delta V({\mathbf x}_\perp) \, \delta V({\mathbf x}'_\perp) \;,
\end{equation} 
with:
\begin{equation}\label{eq:defF}
F({\mathbf x}_\perp ,{\mathbf x}'_\perp) 
\;=\; \frac{1}{16} \, 
\int_{-\infty}^{+\infty} \frac{dk_0}{2\pi} \, 
\Big[
\langle {\mathbf x}_\perp  | ( {\mathbb H}_a )^{-1} | {\mathbf x'}_\perp \rangle
\langle {\mathbf x'}_\perp | ( {\mathbb H}_a )^{-1} | {\mathbf x}_\perp  \rangle
\Big] \;.
\end{equation}

The kernel $F$, which then determines $E_{loc}^{(2)}$ can be evaluated for different 
potentials. The simplest possible case corresponds to $V_a \equiv 0$; in other words, to 
$V_b = \delta V$. Then, $F$ may be calculated exactly, since the inverse
of ${\mathbb H}_a$ is identical to a free scalar field propagator in $2+1$ dimensions,
and the integral can be exactly calculated using standard methods. After some straightforward
steps, one gets the result:
\begin{equation}\label{eq:defFf}
F({\mathbf x}_\perp ,{\mathbf x}'_\perp) 
\;=\; \frac{1}{128} \, \int \frac{d^2{\mathbf k}_\perp}{(2\pi)^2} \, 
e^{i {\mathbf k}_\perp \cdot ({\mathbf x}_\perp - {\mathbf x}'_\perp)}
\; \frac{1}{|{\mathbf k}_\perp|} \;.
\end{equation}
Thus, in terms of the Fourier transform of $\delta V$,
\begin{equation}\label{eq:eloc2f}
E_{loc}^{(2)} \;=\; \frac{1}{256} \, 
\int \frac{d^2{\mathbf k}_\perp}{(2\pi)^2} \;
\frac{|\widetilde{\delta V}({\mathbf k}_\perp)|^2}{|{\mathbf k}_\perp|}\;.
\end{equation} 

Finally, for the case of $V_a$ corresponding to an infinite potential well of width $L$
along the $x_1$ coordinate, which vanishes when $0 < x_1 < L$,
we can write the corresponding kernel as follows:
\begin{align}
F({\mathbf x}_\perp ,{\mathbf x}'_\perp) 
&=\; \frac{1}{4 L^2} \, \int_{-\infty}^{+\infty} \frac{dk_2}{2\pi} \, 
e^{i k_2 (x_2 - x'_2)}\; \sum_{m,n=1}^\infty \, \Big[
I_{mn}(k_2^2)  \nonumber\\  
& \times \,\sin(\frac{m \pi x_1}{L}) \, \sin(\frac{m \pi x'_1}{L}) 
\sin(\frac{n \pi x'_1}{L}) \, \sin(\frac{n \pi x_1}{L}) \Big] \;,
\end{align}
where:
\begin{equation}
I_{mn}(k_2^2) \;=\; \frac{1}{4\pi \sqrt{\Delta_{mn}}} \, 
\log\left[
\frac{(\sqrt{\Delta_{mn}} + k_2^2)^2 - 
\big( (\frac{m \pi}{L})^2 - (\frac{n \pi}{L})^2\big)^2}{(\sqrt{\Delta_{mn}} - k_2^2)^2 - 
\big( (\frac{m \pi}{L})^2 - (\frac{n \pi}{L})^2\big)^2 }
\right] \;,
\end{equation}
and
\begin{equation}
	\Delta_{mn} \;=\; [k_2^2 + (\frac{m \pi}{L})^2 - (\frac{n \pi}{L})^2]^2 + 
4 k_2^2 (\frac{n \pi}{L})^2 \;.
\end{equation}

The results of this section can be extended to more general stepwise Hamiltonians describing stratified media, with value ${\mathbb H}_i$ in the interval
$l_{i-1}<x_d<l_i$,  
\begin{equation}
{\mathbb H}(x_d)=\sum_{i=1}^N {\mathbb H}_i\theta(x_d-l_{i-1})\theta(l_i-x_d)\, ,
\end{equation}
where $l_0=-L_d/2<l_1<l_2<\cdots<l_{N-1}<l_N=L_d/2$. In this case
\begin{equation}
{\mathbb U}(L_d/2,-L_d/2)={\mathbb U}(L_d/2,l_{N-1}){\mathbb U}(l_{N-1},l_{N-2})\cdots{\mathbb U}(l_1,-L_d/2)\, ,
\end{equation}
and each factor in the evolution operator can be evaluated as in Eq.\eqref{eq:umat1}.
\subsection{Magnus expansion for ${\mathbb U}$}
We consider here an approximate method to the calculation of the operator
which appears in the reduced determinant. Since the operator ${\mathbb U}$
satisfies an evolution equation (with a
non-Hermitian `Hamiltonian'), the possibility of using the Magnus expansion \cite{Magnus}
for the calculation of the operator ${\mathbb U}(L_d/2,-L_d/2)$ energy
suggests itself. 
From the equation satisfied by this operator, we have:
\begin{equation}
	{\mathbb U}(L_d/2,-L_d/2) \;=\; \exp\big( \sum_{k=1}^\infty A_n
	\big)
\end{equation}
where 
\begin{align}
	A_1 &= \, \int_{-L_d/2}^{L_d/2} dx_d {\mathcal H}(x_d) \;,\nonumber\\ 
	A_2 &= \, \frac{1}{2} \int_{-L_d/2}^{L_d/2} dx_d
	\int_{-L_d/2}^{x_d} 
	dy_d [{\mathcal H}(x_d),  {\mathcal H}(y_d)] \;, \nonumber\\
	A_3 &= \, \frac{1}{6} \int_{-L_d/2}^{L_d/2} dx_d
	\int_{-L_d/2}^{x_d} 
	dy_d \int_{-L_d/2}^{y_d} dz_d \Big( 
	\big[ {\mathcal H}(x_d),  [ {\mathcal H}(y_d),  {\mathcal H}(z_d)]
	\big] \nonumber\\
	& \;\;\;\;\; \,+\, \big[ {\mathcal H}(z_d),  [ {\mathcal H}(y_d),  {\mathcal
	H}(x_d)] \big] \Big) \;,\nonumber\\
	& \ldots
\end{align}
In this expansion, each term corresponds to a contribution to the exponent of
the operator ${\mathbb U}(L_1/2,-L_1/2)$. We believe that approximation is
convenient here, since the effective action is somehow related to the log of
that operator.  

The $A_1$ term appearing in the expansion above becomes: 
\begin{align}
	A_1 \;&=\; \left( 
	\begin{array}{cc}
		0 & L_d  {\mathbb I} \\
		\int_{-L_d/2}^{L_d/2} dx_d {\mathbb H}(x_d) & 0 
	\end{array}
\right)  \nonumber\\
 &=\; L_d \, \left( 
	\begin{array}{cc}
		0 &  {\mathbb I} \\
		\langle {\mathbb H}\rangle & 0 
	\end{array}
\right) \;.
\end{align}
where we have introduced:
\begin{equation}
\langle {\mathbb H} \rangle \;=\; -\nabla^2_{\mathbf{x}_\perp} +  V_0
({\mathbf x}_\perp) + k_0^2 \;\;,\;\;\; 
V_0({\mathbf x}_\perp) \;\equiv \; \frac{1}{L_d}
\int_{-L_d/2}^{L_d/2} dx_d V_{x_d}({\mathbf x}_\perp) \;.
\end{equation}
To lowest order, the Magnus approximation gives the average Hamiltonian theory~\cite{AHT}, widely used 
in the context of NMR for Hamiltonians with a periodic time dependence.

The next term in the expansion, $A_2$, is given by:
\begin{equation}
	A_2 \;=\; - L_d \,  \frac{1}{2} \int_{-L_d/2}^{L_d/2} dx_d
	\int_{-L_d/2}^{x_d} 
	dy_d \, 
	[ V_{x_d}({\mathbf x}_\perp) - V_{y_d}({\mathbf x}_\perp) ] 
	\,\otimes
	\left( \begin{array}{cc}
1  &  0	\\ 
0 & -1 
	\end{array}
\right)
\end{equation}
or, introducing $V_1({\mathbf x}_\perp) \equiv \frac{1}{L_d} \,\int_{-L_d/2}^{L_d/2}
	dx_d \, x_d \, V_{x_d}({\mathbf x}_\perp)$:
\begin{equation}
	A_2 \;=\; - L_d \,  V_1 ({\mathbf x}_\perp)
	\otimes
	\left( \begin{array}{cc}
1  &  0	\\ 
0 & -1 
	\end{array}
\right) \;.
\end{equation}

We conclude this section by deriving a more explicit expression for the
energy when one takes into account the terms up to the second order, $A_1$
and $A_2$ above; namely, when we approximate:
\begin{equation}
	{\mathbb U}(L_1/2,-L_1/2) \;\simeq\; \exp\big( A_1 + A_2) \;.
\end{equation}
We see, after some lengthy algebra, that:
\begin{equation}
v_1^T \exp\big( A_1 + A_2) v_2 \;=\; \frac{\sinh(L_1 \sqrt{\mathbb
H}_2)}{\sqrt{\mathbb H}_2} \;,
\end{equation}
with:
\begin{equation}
{\mathbb H}_2 \;=\; -\nabla^2_{\mathbf{x}_\perp} +  V_0
({\mathbf x}_\perp) + \big( V_1({\mathbf x}_\perp) \big)^2 + k_0^2 \;, 
\end{equation}
namely, the energy may be written in terms of the Casimir energy for a
system which has an `effective potential' $V_{eff}$ independent of $x_1$, 
but modified by the dependence on $x_1$ of the real, original potential:
\begin{equation}
	V_{eff}({\mathbf x}_\perp) \;=\; V_0({\mathbf x}_\perp) \,+\,
	\big( V_1({\mathbf x}_\perp) \big)^2 \;.
\end{equation}

A possible example where one can apply in a rather natural way the Magnus
expansion, corresponds to the case of a potential modulated in strength. 
Namely, we consider here a situation such that the potential $V_{x_d}({\mathbf
x}_\perp) = g(x_d) \, V({\mathbf x}_\perp)$. 
Thus, $V_0({\mathbf x}_\perp) = g_0 V({\mathbf x}_\perp)$, and 
$V_1({\mathbf x}_\perp) = g_1 V({\mathbf x}_\perp)$, with:
\begin{align}
	g_0 \;&\equiv \; \frac{1}{L_d} \int_{-L_d/2}^{L_d/2} dx_d \, g(x_d)
	\nonumber\\
	g_1 \;&\equiv \; \frac{1}{L_d} \int_{-L_d/2}^{L_d/2} dx_d
x_d \, g(x_d) \;.	
\end{align}

Thus
\begin{equation}
	V_{eff}({\mathbf x}_\perp) \;=\; g_0 \,V({\mathbf x}_\perp) \,+\,
	g_1^2 \, 
	\big( V({\mathbf x}_\perp) \big)^2 \;.
\end{equation}

The vacuum energy is then the one of a system with an effective potential
as the one above, depending on $d-1$ spatial dimensions. It is interesting to remark that, if the leading correction is nonvanishing ($g_0\neq 0$), then there is a choice of the origin of coordinates such that $g_1=0$. 

Let us consider the $3+1$ dimensional example of mirrors represented by square potentials along
a coordinate $x_2$, with a strength modulated along $x_d = x_3$. Since in this
case the effective potential is also square, we may apply
Lifshitz~\cite{Lifshitz:1956zz} formula for the Casimir energy density. Indeed, it has been 
shown~\cite{CcapaTtira:2011ga} that for two thick and homogeneous slabs modeled by square potentials of amplitude $V^{(1)}$ and $V^{(2)}$, the Casimir energy density reads
\begin{equation}\label{Lifshitz}
{\mathcal E} \,=\, -\frac{1}{2} \int \frac{d^dk}{(2\pi)^d} \, 
 \log \Big[ 1 -
\frac{ V^{(1)}  V^{(2)}}{(k+\bar\Omega^{(1)})^2(k+\bar\Omega^{(2)})^2}
e^{- 2k l} \Big] \, ,
\end{equation}
where  $l$ is the distance between slabs and $\bar\Omega^{(i)}=\sqrt{k^2+V^{(i)}}$. 
For semi infinite slabs, the potential reads
\begin{equation}\label{squareV}
V(x_2)=V^{(1)}\theta(-x_2)+V^{(2)}\theta(x_2-l)\, .
\end{equation}
In physical terms, 
\begin{equation}
r^{(i)}= \frac{ V^{(i)}}{(k+\bar\Omega^{(i)})^2}
\end{equation}
are the reflection coefficients of each slab. In a more realistic model involving the electromagnetic field the reflection coefficients will depend on the permittivity and permeability of each material. 

Now consider a generalized problem in which the properties of the materials 
depend on a coordinate parallel to the mirrors, $x_3$, that is, the materials are 
not homogeneous in this direction. This problem can be modeled by a square potential at each $x_1$, i.e. $V_{x_3}(x_2)=g(x_3)V(x_2)$. The lowest order Magnus 
approximation 
to the Casimir energy can be obtained from Lifshitz formula replacing
$V^{(i)}\to V^{(i)}_{eff}=g_0 V^{(i)}$. This corresponds
to the average of the material properties along the non-homogeneous direction. From a physical point of view, 
this approximation should be accurate when the scale of variation of the function $g(x)$ is much smaller than $l$. In the opposite
limit, we expect the Casimir energy to be well approximated by the proximity force approximation (PFA). Assuming
that $g(x)$ is piecewise constant, taking the value $g^{(a)}$ over a length $d_a$, the PFA to the Casimir energy
density  turns out to be
equal to the average of the Casimir energies for each piece
\begin{equation}
 {\mathcal E}_{PFA}= \frac{1}{L_1} \sum_i{\mathcal E}^{(a)}d_a \, .
 \end{equation}
where  ${\mathcal E}^{(a)}$ is the Casimir energy corresponding to the potentials $g^{(a)} V^{(i)}$. Therefore, the Magnus approximation
can be thought of as complementary to the PFA: while the former averages the properties of the materials, the latter averages the 
 Casimir energies.

\section{Conclusions}\label{sec:concu}
We have presented a possible way to study fluctuation determinants
for a real scalar field in $d+1$ dimensions, by applying a generalized
version of G-Y  theorem. The outcome is that the original object is now
transformed into the determinant of a (function of a) lower-dimensional
operator. 

This approach, which certainly could me applied to other contexts, has been
tested for the Casimir effect. We have shown that it allows for the
introduction of a novel approximation scheme. Indeed, since the reduction
process involves a kind of evolution operator for a `time'-dependent
Hamiltonian, we have been able to apply the Magnus expansion to deal with
the calculation of that object, and therefore of the Casimir energy. We
evaluated the first non-trivial terms in that expansion within the context
of a concrete example, showing that it amounts to an alternative approximation
scheme.

Non-perturbative analytical calculations are usually much harder to obtain.
We have considered a particular example, involving two half-infinite
non-homogeneous media, where non-perturbative results for the vacuum energy
could be exactly calculated.  In particular, we have shown that there is a
contribution which can naturally be attributed to the interface between the
media. That contribution has been evaluated explicitly in perturbation
theory, under some extra the assumptions regarding the properties of the two media.

Finally, we wish to stress that the result of the reduction process could
 be useful when using numerical approaches to the determination of the
 vacuum energy.


\end{document}